\documentstyle[11pt]{article}
\title{\bf On representation of the $t-J$ model via spin-charge variables}
\author{ \vspace*{2mm}
{\bf Evgueni Kochetov and Vladimir Yarunin} \\
\small \it Bogoliubov Laboratory of Theoretical Physics,\\
\small \it Joint Institute for Nuclear Research, 141980 Dubna, Russia}
\begin{document}
\date{}
\maketitle
\begin{abstract}
We show that the $t-J$ Hamiltonian is not in general reduced to
$H_{t-J}=H(\vec S,f)$, where $\vec S$ and $f$ stand for independent
$([\vec S,f]=0)$ $SU(2)$
(spin) generators and spinless fermionic (hole) field, respectively.
The proof is based upon an identification of the Hubbard operators with the
generators of the $su(2|1)$ superalgebra in the degenerate fundamental
representation and ensuing $SU(2|1)$ path integral representation
of the partition function $Z_{t-J}$.
\end{abstract}

e-mail:kochetov@thsun1.jinr.dubna.su, fax number:(7)(096)(21)65084

\vspace{1cm}
{\Large \bf I. Introduction}
\vspace{1cm}

It is by now widely accepted that the $t-J$ model,
that is, the one-band Hubbard model in the large $U$-limit,
provides an adequate basis for the discussion of the essential physics
for layered cooper oxide compounds \cite{and,zhang}.
An accurate description of the properties of charge carriers
in high-temperature superconductors arising from their interaction with
the spin of the $Cu$ atoms seems to be crucial for the understanding of
superconductivity in these materials. Since there occurs a strong coupling
between charge and spin degrees of freedom \cite{kane},
the problem of a proper separation of spin and charge degrees
of freedom in the $t-J$ model
is of importance in order to get an insight into an interplay of magnetic
and charge properties of relevant systems.

A popular approach has so far been that to introduce
chargeless spinon and spinless
holon operators in the framework of the slave fermion or slave boson
method. The drawback of this approach is, however, that a certain local
(at every lattice site) constraint on spinon and holon operators is to be
imposed in order to ensure the single occupancy of the electrons.
Although the spinon and holon degrees of freedom
are separated on the mean field level, they are strongly coupled by the gauge
field associated with fluctuations around a mean field \cite{lee}.
It has been
recognized that while the rigorous imposition of the constraint seems
to pose a problem, its averaging, e.g., mean field treatment
results in a large error.

It seems therefore desirable to attempt to explicitly formulate the $t-J$
model in terms of independent local spin and holon operators so that
no constraint would be necessary.
Some recent developments point to the possibility to attain the goal
starting from a new kind of spin-fermion representation
for the Hubbard operators \cite{yu,wang}. Although this representation
agrees with the required commutation relations for the Hubbard
operators, it implies that the
original Hilbert space is to be enlarged, and as a result
a certain constraint
seems to be necessary anyway to get rid of unphysical degrees of freedom.
Besides, the enlargement is not entirely fixed
in the scope of this approach.

In the present paper we show that the $t-J$ Hamiltonian is not in
general reduced to a polynomial function of independent $SU(2)$
(spin) and fermion (hole) variables, though that happens in the so-called
linear spin wave approximation.
We address the problem from the general point of view,
by considering the Hubbard operators as the generators of the $su(2|1)$
superalgebra and employing the ensuing $SU(2|1)$ path integral representation
for the partition function. The $SU(2|1)$ supersymmetry happens to be
the largest symmetry that underlies the $t-J$ model \cite{wig}.
In essence, our approach is nothing but the geometric quantization
(also called {\it the coherent state method})
for quantum mechanics associated
with a semisimple Lie algebra \cite{perelomov,gradechi,topi}.
It provides an effective, in the sense it requires a minimal
set of variables,
description of a system with a Hamiltonian that can be embedded into a given
Lie (super)algebra. As an example, we may refer to the $SU(2)$
path-integral representation of a partition function
that has recently been employed
to formulate a non-operator mean-field diagrammatic technique for the
Heisenberg model \cite{kochetov2}. Path integral associated with $su(2|1)$
supercoherent states has proved to be helpful in order to justify
the adiabatic approximation in the periodic Anderson model in the
large $U$-limit \cite{kochetov5}.

The proof of the statement given in the abstract is quite simple,
though it requires that
some necessary notation is to be introduced first,
so that a bulk of the paper serves to that purpose.
Section II, as well as Appendices A, B and C are necessary to make
the proof given in section III quite transparent and plain. Section IV
explains an exception that is provided by the linear spin wave theory.
Section V contains some comments on earlier results and concluding remarks.

\vspace{1cm}
{\Large \bf II. $t-J$ model}
\vspace{1cm}

We start by expressing the $t-J$ model in terms of the Hubbard operators
\cite{hubbard} $X^{\sigma 0}_i$, defined as
$$X^{\sigma 0}_i=c^{\dag}_{i\sigma}(1-n_{i,-\sigma}),$$
where $c_{i\sigma}$ is the annihilation operator of an electron at site $i$
with spin $\sigma=\pm$, and $n_{i\sigma}\equiv c^{\dag}_{i\sigma}c_{i\sigma}$.
In terms of these, the $t-J$ Hamiltonians becomes
\begin{eqnarray}
H_{t-J}&=&-t\sum_{ij\sigma}X^{\sigma 0}_iX^{0\sigma}_j +
J\sum_{ij}\vec Q_i\vec Q_j,
\label{eq:3.1}\end{eqnarray}
where the electron spin operator
$$\vec Q_i=\frac{1}{2}\sum_{\sigma\sigma'}X^{\sigma0}_i
\vec{\tau}_{\sigma\sigma'}X^{0\sigma'}_i,$$
and $\vec{\tau}=(\tau^1,\tau^2,\tau^3)$ are the Pauli matrices.

For further convenience, though it is not necessary to prove
our statement, we perform a $\pi$-rotation of the spins on the $B$-sublattice
which leads to the changes
$$X^{0\sigma}_i\to X^{0-\sigma}_i,\quad Q^{\pm}_i\to Q^{\mp}_i,\quad
Q^z_i\to -Q^z_i,\quad i\in B.$$
Hence from now on the spin background is effectively a ferromagnetic one
and one should not distinguish between sublattices anymore.
The original Hamiltonian (\ref{eq:3.1}) is then converted into
\begin{eqnarray}
H_{t-J}=-t\sum_{ij}\left(X^{+0}_iX^{0-}_j+X^{-0}_iX^{0+}_j
\right)
+J\sum_{ij}\left[-Q^z_iQ^z_j+\frac{1}{2}
(Q^{+}_iQ^{+}_j +Q^{-}_iQ^{-}_j)\right]
\label{eq:3.2}\end{eqnarray}

$X^{\sigma 0}$ projects the electron operator into the single-occupation
state and in the basis $\{|0\rangle,|\sigma\rangle\}$ takes the form
\begin{eqnarray}
X^{\sigma 0} =|\sigma\rangle\langle 0|,\quad X^{\sigma\sigma'}=
|\sigma\rangle\langle\sigma'|,
\label{eq:3.3}\end{eqnarray}
where $|0\rangle$ stands for a doped site (hole) and $|\sigma\rangle$
for the state having an electron occupied with spin $\sigma$.
It is clear that there are eight linearly independent operators since
\begin{eqnarray}
X^{00}+\sum_{\sigma}X^{\sigma\sigma}=I,
\label{eq:3.4}\end{eqnarray}
$X^{\sigma 0}$ appearing as a fermionic operator, whereas $X^{\sigma\sigma'}$
correspond to bosonic degrees of freedom.
In fact, representation (\ref{eq:3.3}) means that the $X$-operators
are closed into the $u(2|1)$ superalgebra, which in view of (\ref{eq:3.4})
is reduced to the eight-dimensional $su(2|1)$ superalgebra.
The latter is generated by even generators $\{B,Q_3,Q_{+},Q_{-}\}$ and
the odd ones $\{W_{+},W_{-},V_{+},V_{-}\}$ and the associated coherent
state in the so-called $(q,q)$ representation (see Appendix A) reads
\begin{eqnarray}
|z,\xi\rangle=(1+|z|^2+\bar\xi\xi)^{-q}e^{-\xi V_{-}+z Q_{-}}|q,q,q\rangle,
\label{eq:2.3}\end{eqnarray}
where $|b,q,q_3\rangle$ stands for a eigenvector
of the operators
$B,{\vec Q}^2$ and $Q_3$, respectively,
and
the variables $z$ and $\xi$ parametrize the super-two-sphere
$SU(2|1)/U(1|1)=S^{(2|2)}$, the $N=2$ supersymmetric extension of the
two-sphere $S^2$ (for some details concerning a definition of $S^{2|2}$
see Appendix C).

Resolution of unity in the $(q,q)$ representation space holds
$$\int|z,\xi\rangle\langle z,\xi|d\mu_{SU(2|1)}(z,\xi)=I$$
$$=\sum_{-q}^{q}|q;q,m\rangle\langle q;q,m|+
\sum_{-(q-1/2)}^{q-1/2}|q+1/2;q-1/2,m\rangle\langle q+1/2;q-1/2,m|,$$
provided
\begin{eqnarray}
d\mu_{SU(2|1)}=\frac{d\bar zdz}{2\pi i}
\frac{d\bar\xi d\xi}{1+|z|^2+\bar\xi\xi}.
\label{eq:2.4}\end{eqnarray}

Evaluating a partition function in the $\{|z,\xi\rangle\}$ basis results
eventually in the $SU(2|1)$ path integral representation \cite{kochetov4}
\begin{equation}
tr \,\exp[-\beta H]\equiv Z_{SU(2|1)}=
\int\limits_{z(0)=z_(\beta)}^{\xi(0)=-\xi(\beta)}
D\mu_{SU(2|1)}(z,\xi)\exp\left[{\cal A}\right],
\label{eq:2.5}
\end{equation}
where $D\mu_{SU(2|1)}(z,\xi)$ stands for an infinite pointwise
product of the $SU(2|1)$
invariant measures (\ref{eq:2.4}) and the classical action on $S^{2|2}$
with a Hamiltonian function $H^{cl}=\langle z,\xi|H|z,\xi\rangle$ reads
\begin{eqnarray}
{\cal A}=q\int_0^{\beta}\frac{\dot{\bar z}z-\bar z\dot z
+\dot{\bar\xi}\xi-\bar\xi
\dot\xi}{1+|z|^2+\bar\xi\xi}dt-\int_0^{\beta}H^{cl}(z,\xi)dt.
\label{eq:2.6}\end{eqnarray}
A few important definitions concerning the notion of an integration
on supermanifols are
given in Appendix C.

To explicitly evaluate $H^{cl}$ one needs the $SU(2|1)$ covariant symbols
of the generators. These are found to be
$(A^{cl}\equiv\langle z,\xi|A|z,\xi\rangle)$:
\begin{eqnarray}
Q^{cl}_3&=&-q(1-|z|^2)w,\quad (Q^{+})^{cl}=2qzw, \quad (Q^{-})^{cl}=2q\bar zw,\nonumber\\
B^{cl}&=&q(1+|z|^2+2\bar\xi\xi)w,\quad (V^{+})^{cl}=-2q z\bar\xi w, \quad
(V^{-})^{cl}=2q\bar\xi w,\nonumber\\
(W^{+})^{cl}&=&-2q\xi w,\quad  (W^{-})^{cl}=-2q\bar z\xi w,\quad
w=(1+|z|^2+\bar\xi\xi)^{-1}.
\label{eq:2.7}\end{eqnarray}

Turning back to the $t-J$ model one notices
that the algebra of the $X$-operators can explicitly be identified
with the degenerate $(1/2,1/2)$ representation of $su(2|1)$
in the following way,
$$Q_3=\frac{1}{2}(X^{++}-X^{--}),\quad Q_{+}=X^{+-},\quad Q_{-}=X^{-+},
\quad B=\frac{1}{2}(X^{++}+X^{--})+X^{00}$$
and
$$ V_{+}=X^{0-},\quad V_{-}=-X^{0+},\quad W_{+}=X^{+0},\quad
W_{-}=X^{-0},$$
the even (bosonic) states $|1/2,1/2,1/2\rangle$ and $|1/2,1/2,-1/2\rangle$
being identified with the spin up and spin down states, $|+\rangle$ and
$|-\rangle$ , respectively, whereas the odd (fermionc) state $|1,0,0\rangle$
with the doped state $|0\rangle$. Dimension of this representation is equal
to $3$ as should be. It is also clear that Eq. (\ref{eq:a.2}) holds true and
hence we have explicitly identified the algebra of the Hubbard operators
with the degenerate fundamental $(3\times 3)$ representation of the $su(2|1)$
superalgebra.

It is worth mentioning that $su(2|1)$ gives rise in a natural way
to the slave fermion (slave boson) representation for the Hubbard operators.
The latter appears as
the so-called oscillator representation of the $su(2|1)$ algebra \cite{bars}.
For instance, let $X_{\lambda\lambda'}, \lambda,\lambda'=1,2,3$ be a matrix
corresponding to the operator $X$ in the $(1/2,1/2)$
representation. Consider a composite creation operator
$d^{\dag}=(a^{\dag},b^{\dag},f^{\dag})$, where $a$ and $b$ stand for bosonic
fields and $f$ for a fermionic one. Then, the slave fermion representation
reads
$$X=\sum_{\lambda\lambda'}d^{\dag}_{\lambda}X_{\lambda\lambda'}d_{\lambda'},$$
\begin{equation}
\sum_{\lambda}d^{\dag}_{\lambda}d_{\lambda}=a^{\dag}a
+b^{\dag}b+f^{\dag}f=1,
\label{eq:2.77}\end{equation}
where the last line is the completeness relation ({\ref{eq:3.4}).
In fact, this is nothing but a linear Casimir operator
of $u(2|1)$ whose eigenvalue fixes a representation.
The lowest
possible value taken by the rhs and equal to $1$
corresponds to the lowest possible
dimension of the reprsentation space.

The $su(2|1)$ algebraic approach provides also a possible
generalization of the
standard $t-J$ Hamiltonian to include particles with spin higher than $1/2$,
which is necessary to properly formulate a $1/s$ expansion.
One possibility might be to consider spin $s$ electrons, which would
correspond to the fundamental representation of the $su(2s+1|1)$ superalgebra
instead of $su(2|1)$. An alternative procedure, since we are really interested
in $s=1/2$, is to interpret the holes to be sites which have spin $s-1/2$
\cite{kane}, so that the sites without a "hole" acquire spin $(s-1/2)+1/2=s$.
The latter possibility
amounts to considering the $(q=s,q=s)$
representation of $su(2|1)$ rather than the $(q=1/2,q=1/2)$ fundamental one.
The hole space is then identified with the set
$$\{|q+1/2,q-1/2,m\rangle,\quad -q+1/2\leq m\leq q-1/2\}$$
whereas the "holeless" spin excitations form the set
$$\{|q,q,m\rangle,\quad -q\leq m\leq q\}.$$
This remark clarifies the physical meaning of the representation index $q$.

\vspace{1cm}
{\Large \bf III. $SU(2|1)$ path integral for the $t-J$ model}
\vspace{1cm}

With the necessary background displayed above, one easily arrives at the
$SU(2|1)$ path integral representation for the partition function
$$Z_{t-J}=tr\,e^{-\beta H_{t-J}}.$$
The result is
\begin{eqnarray}
Z_{t-J}=\int_{S^{2|2}}\prod_j
D\mu^{(j)}_{SU(2|1)}\exp[{\cal A}_{t-J}],\quad
z_j(0)=z_j(\beta),\,\xi_j(0)=-\xi_j(\beta),
\label{eq:4.1}\end{eqnarray}
where
\begin{eqnarray}
{\cal A}_{t-J}=q\sum_j\int_0^{\beta}\frac{\dot{\bar z_j}z_j-\bar z_j\dot z_j
+\dot{\bar\xi_j}\xi_j-\bar\xi_j\dot\xi_j}{1+|z_j|^2+\bar\xi_j\xi_j}dt-
\int_0^{\beta}H^{cl}_{t-J}dt
\label{eq:4.2}\end{eqnarray}
The first term of the action is purely geometric and reflects the
structure of $S^{2|2}$ while the second is of a dynamical origin
and in view of (\ref{eq:2.7}) is found to be
\begin{eqnarray}
H^{cl}_{t-J}&=&-t(2q)^2\sum_{ij}\frac{[\xi_i\bar\xi_jz_j +
\bar z_i\xi_i\bar\xi_j]}
{(1+|z_i|^2)(1+|z_j|^2)}
\nonumber\\
&&+Jq^2\sum_{ij}\frac{\left[-(1-|z_i|^2)(1-|z_j|^2)+2(z_iz_j+
\bar z_i\bar z_j)\right]}
{(1+|z_i|^2+\bar\xi_i\xi_i)(1+|z_j|^2+\bar\xi_j\xi_j)}.
\label{eq:4.3}\end{eqnarray}
To avoid an accumulation of indices, we will often drop the
lattice site indication whenever no confusion is possible.

Representation (\ref{eq:4.1}-\ref{eq:4.3}) is the point we will start from to
prove the main statement of the paper. We will proceed as follows.
Suppose we are given a Hamiltonian to be a function (polynomial)
of the spin generators $\vec S$ and spinless fermionic fields $f,\,f^{\dag}$
(for the notation see Appendix B),
$$H_{\vec S-f}=H(\vec S,f).$$
Then it follows (see Appendix B) that
\begin{eqnarray}
Z_{\vec S-f}=\int_{S^2}D\mu_{SU(2)}\int D\mu_F\exp[{\cal A}_{\vec S-f}],
\label{eq:4.4}\end{eqnarray}
where
\begin{eqnarray}
{\cal A}_{\vec S-f}=s\int_0^{\beta}
\frac{\dot{\bar z}z-\bar z\dot z}{1+|z|^2}dt+
\frac{1}{2}\int_0^{\beta}(\dot{\bar\xi}\xi-\bar\xi\dot\xi)dt-
\int_0^{\beta}H^{cl}_{\vec S-f}dt,
\label{eq:4.5}\end{eqnarray}
and
$$H^{cl}_{\vec S-f} =\langle z|_{SU(2)}\langle\xi|_F\, H_{\vec S-f}\,
|\xi\rangle_F |z\rangle_{SU(2)}.$$
The first two terms in ${\cal A}_{\vec S-f}$ are of a
geometric origin, as well.

Let us now compare Eqs. (\ref{eq:4.1}) and (\ref{eq:4.4}). If it were possible
by a change of variables to bring somehow the first equation
to the form of the
second one, it would mean that the $t-J$ Hamiltonian can be reduced to
a certain $H_{\vec S-f}$. If one failed to do this, it
would not in general mean that $H_{t-J}$ coincides with no $H_{\vec S-f}$.
It may just mean that we have failed to find a proper transformation that
would result in a decomposition
\begin{eqnarray}
D\mu_{SU(2|1)}&\to& D\mu_{SU(2)}\,D\mu_F\nonumber \\
q\int_0^{\beta}\frac{\dot{\bar z_j}z_j-\bar z_j\dot z_j
+\dot{\bar\xi_j}\xi_j-\bar\xi_j\dot\xi_j}{1+|z_j|^2+\bar\xi_j\xi_j}dt
&\to& s\int_0^{\beta}
\frac{\dot{\bar z}z-\bar z\dot z}{1+|z|^2}dt+
\frac{1}{2}\int_0^{\beta}(\dot{\bar\xi}\xi-\bar\xi\dot\xi)dt,
\label{eq:4.6}\end{eqnarray}
where $s=s(q)$.
This decomposition is necessary to arrive at in view of Eq. (\ref{eq:4.5}).
It should be recognized that both lines of Eq. (\ref{eq:4.6})
are to be fulfilled {\it simultaneously}. It might also be possible that
a corresponding change of variables does not exist in principle, though both
Eqs. (\ref{eq:4.1}) and (\ref{eq:4.4}) may contain the same physical
information, which would in turn mean that the path integral approach
fails to provide a definite answer,
which in itself is very unlikely.

Once, on the other hand, one has succeeded with Eq. (\ref{eq:4.6})
the next step to take would be to look at a form the $t-J$ Hamiltonian
is transformed to in accordance with
(\ref{eq:4.6}). If the latter coincided with
the covariant symbol of a certain $H_{\vec S-f}$, then one could conclude
\begin{equation}
H_{t-J} = H_{\vec S-f}.
\label{eq:4.eq}\end{equation}
Note that there is one-to-one correspondence between $H_{\vec S-f}$ and its
covariant symbol \cite{berezin2}.
In case the transformed $\tilde {H}^{cl}_{t-J}$
cannot be identified with the covariant symbol of any $H_{\vec S-f}$,
Eq. (\ref{eq:4.eq}) does not hold. It is just the case for the
$t-J$ model.

To prove that, let us in Eq. (\ref{eq:4.1}) make two successive
changes of variables:
\begin{equation}
z_i\to z_i\sqrt{1+\bar\xi_i\xi_i},\quad \mbox{ and }\quad
\xi_i\to\xi_i\sqrt{\frac{1+|z_i|^2}{2q}},
\label{eq:4.tr}\end{equation}
which results in
\begin{eqnarray}
Z_{t-J}&\to& Z_{t-J}= \int_{S^2}\prod_j
D\mu^{(j)}_{SU(2)}\int\prod_jD\mu^{(j)}_F\exp[\tilde{\cal A}_{t-J}]\nonumber\\
\tilde{\cal A}_{t-J}&=&q\sum_j\int_0^{\beta}
\frac{\dot{\bar z_j}z_j-\bar z_j\dot z_j}{1+|z_j|^2}dt+
\frac{1}{2}\sum_j\int_0^{\beta}(\dot{\bar\xi_j}\xi_j-\bar\xi_j\dot\xi_j)dt-
\int_0^{\beta}\tilde{H}^{cl}_{t-J}dt.
\label{eq:4.7}\end{eqnarray}
The important point concerning this representation is that $z(t)$ and
$\bar z(t)$ can be considered to take values in $S^2$ in accordance with
Eq. (\ref{eq:b.3}) of Appendix C. Hence, we have succeeded in converting
the $SU(2|1)$ integral (\ref{eq:4.1}) into the $SU(2)$ and the purely
fermionic ones. It is also seen from (\ref{eq:4.7}) and (\ref{eq:a1},
\ref{eq:a2}) that $s(q)=q$, that is, we have arrived at the $SU(2)$
representation with spin $s=q$. To summarize, we have separated spin
and charge variables at {\it the kinematical} level. In other words,
a possibility of the spin-charge variable
separation for any model expressible in terms of the Hubbard operators
depends solely on an explicit form of the Hamiltonian.

Let us write down the function $\tilde{H}^{cl}_{t-J}$
explicitly:
\begin{eqnarray}
\tilde{H}^{cl}_{t-J} &=& -t(2q)\sum_{ij}\frac{\xi_i\bar\xi_jz_j
+\bar z_i\xi_i\bar\xi_j}
{\sqrt{(1+|z_i|^2)(1+|z_j|^2)}}\nonumber\\
&& J\sum_{ij}\left[-\left(-q\frac{1-|z_i|^2}{1+|z_i|^2}
+\frac{\bar\xi_i\xi_i}{2}\right)\left(-q\frac{1-|z_j|^2}{1+|z_j|^2}
+\frac{\bar\xi_j\xi_j}{2}\right)\right.\nonumber\\
&&\left.+\frac{1}{2}\left(\frac{2qz_i}{1+|z_i|^2}-\frac{z_i\bar\xi_i\xi_i}
{2}\right)\left(\frac{2qz_j}{1+|z_j|^2}-\frac{z_j\bar\xi_j\xi_j}{2}\right)
+h.c.\right]
\label{eq:4.8}\end{eqnarray}
What conclusion can be drawn from this representation? Some terms can
be viewed as covariant symbols of spin-fermion interaction operators.
For instance, the second line in Eq. (\ref{eq:4.8}) is simply a symbol
of the operator (see Appendix B)
$$-J\sum_{ij}(S^z_i-n_i/2)(S^z_j-n_j/2),$$ where $n_i=f^{\dag}_if_i$ is
the hole number operator. Besides, it is clear that
$$\frac{2qz_i}{1+|z_i|^2}=\langle z_i|S_i^{+}|z_i\rangle,$$
etc. On the other hand, there is no a polynomial function $f(\vec S)$ with
the property
$$\langle z|f|z\rangle=z,$$
as well as there is no such $f(\vec S)$ that would give rise
to the square roots in the $t$-dependent term.
It is also obvious that the change of the variables (\ref{eq:a5})
is of no use in order to get rid of the unwanted terms,
and we finally conclude, there is no
$H_{\vec S-f}$ such that Eq. (\ref{eq:4.eq}) would hold.

To complete the proof, two remarks are in order. First, we compare in fact
classical actions (Lagrangians) (\ref{eq:4.2}) and (\ref{eq:4.5}) as well as
related integration measures (invariant volume elements) rather than
partition functions $Z_{t-J}$ and $Z_{\vec S-f}$. This implies that a path
integral does not seem to be indispensable for the above consideration.
Classical action can be obtained by standard methods.
Namely, given a (super)coherent state $|z\rangle$ with $z$ being a set of
supercoordinates,
one can obtain a corresponding action
${\cal A}$ with the help of equation
$$ {\cal A}=i\int \langle z|\frac{d}{dt}-H|z\rangle dt.$$
To evaluate this explicitly, representation
$$d/dt=\dot z\partial_z+\dot{\bar z}\partial_{\bar z}$$
is to be used.
We prefer, however, to employ the path-integral
formalism since it provides the most simple consideration.

Second, it is not sufficient for our purposes to merely compare
Hamiltonians $H_{t-J}^{cl}$ and $H_{\vec S-f}^{cl}$. The point is that
canonical equations of motion
that follow from the Hamiltonian action
principle $\delta{\cal A}=0$ and read
$$\dot z=\{H^{cl},z\},$$
depend on both classical Hamiltonian and underlying geometry.
Here $\{,\}$ stands for the Poisson brackets which
involve different symplectic
two forms $\omega$ for different manifolds. Actually, the form $\omega$
defines a kinetic term in an action which can be written in the form
$i\int \theta$ where $d\theta=\omega$.
That is why it is
necessary to compare either Lagrangians or Hamiltonians plus corresponding
two-forms (invariant volume elements).

\vspace{1cm}
{\Large \bf IV. Linear spin-wave approximation}\\
\vspace{1cm}

As is shown in the preceding section it is in general impossible to reduce
the $t-J$ hamiltonian to that of a spin-fermion interaction.
Now we demonstrate how this can be achieved in the so-called
linear spin-wave (LSW) approximation \cite{anderson,kubo},
which effectively corresponds to small transverse fluctuations of a spin
around the $z$ axis. As is seen from  Eq. (\ref{eq:a3}), this mathematically
means $|z|^2\ll 1$ (in fact, $|z|^2/2q\ll 1$).
The LSW theory
has been successfully exploited in the $t-J$ model, see the paper
\cite{horsch} and references therein.

In the path-integral language the LWS approximation consists in converting
the $SU(2)$ path integral representation
(\ref{eq:a1},\ref{eq:a2}) under the condition $|z|^2\ll 1$
into the bosonic one.
To proceed, one should expand the action (\ref{eq:a2}) up to the second
order in $z,\bar z$ and perform a change $z\to z/\sqrt{2q}$, the latter
being needed to recover in the action the "flat" kinetic term:
$$ \frac{1}{2}\int_0^{\beta}(\dot{\bar z}z-\bar z\dot z)dt.$$
With all this having been performed, Eq. (\ref{eq:4.1}) becomes
\begin{eqnarray}
Z_{t-J}\to Z^{LSW}_{t-J}=\int\,\prod_jD\mu^{(j)}_B(z)
D\mu^{(j)}_F(\xi)\,\exp[{\cal A}^{LSW}_{t-J}],
\label{eq:5.1}\end{eqnarray}
where
\begin{eqnarray}
{\cal A}^{LSW}_{t-J}=
\frac{1}{2}\sum_i\int_0^{\beta}
(\dot{\bar z_i}z_i-\bar z_i\dot z_i)dt+
\frac{1}{2}\sum_i\int_0^{\beta}(\dot{\bar\xi_i}\xi_i-\bar\xi_i\dot\xi_i)dt-
\int_0^{\beta}H^{cl}_{LSW}dt,
\label{eq:5.2}\end{eqnarray}
and $D\mu_B(z)=D\bar zDz.$
On the other hand,
\begin{eqnarray}
&&H^{cl}_{t-J}\to H^{cl}_{LSW}=-t\sqrt{2q}\sum_{ij}(\xi_i\bar\xi_jz_j
+\bar z_i\xi_i\bar\xi_j)\nonumber\\
&&+Jq\sum_{ij}\left[\bar z_iz_i(1-\frac{\bar\xi_j\xi_j}{2q})
+\bar z_jz_j(1-\frac{\bar\xi_i\xi_i}{2q})-q(1-\frac{\bar\xi_j\xi_j}{2q})
(1-\frac{\bar\xi_i\xi_i}{2q})\right.\nonumber\\
&&\left. +z_iz_j(1-\frac{\bar\xi_i\xi_i}{4q})(1-\frac{\bar\xi_j\xi_j}{4q})
+\bar z_i\bar z_j(1-\frac{\bar\xi_i\xi_i}{4q})(1-\frac{\bar\xi_j\xi_j}{4q})
\right],
\label{eq:5.3}\end{eqnarray}
which may be regarded as the covariant symbol in the LSW limit
of the operator
\begin{eqnarray}
H^{LSW}_{t-J}&=&-t\sum_{ij}(f_if^{\dag}_jS^{+}_j+f_if^{\dag}_jS^{-}_i)
+J\sum_{ij}\left[-(S^z_i-n_i/2)(S^z_j-n_j/2)\right.\nonumber\\
&&\left.+\frac{1}{2}S^{+}_iS^{+}_j (1-\frac{n_i}{4q})(1-\frac{n_j}{4q})
+\frac{1}{2}S^{-}_iS^{-}_j (1-\frac{n_i}{4q})(1-\frac{n_j}{4q})\right].
\label{eq:5.4}\end{eqnarray}
Here $f^{\dag}$ and $f$ stand for spinless hole operators, $n=f^{\dag}f$,
while $\vec S$ describes a local spin, with $[\vec S,f]=0$.
It is important to recognized that when deriving (\ref{eq:5.4})
we have not been forced to impose any constraints (cf. Ref. \cite{horsch}),
since we have started off
with Eq. (\ref{eq:3.1}) that automatically implies no double occupied
configurations and made no algebraic identifications  of the Hubbard
operators with spin-fermion bilinears (cf. Ref. \cite{yu}).
Representation (\ref{eq:5.3}-\ref{eq:5.4}) coincides with that of Ref.
\cite{horsch}, provided the mean-field approximation $n_i=\delta\ll 1$, where
$\delta$ is the concentration of holes, is used.

\vspace{1cm}
{\Large \bf V. Comments and conclusion}
\vspace{1cm}

As was already mentioned, there exist papers where some explicit
representations of the form
$$H_{t-J}=H(\vec S,f)$$
have been obtained, e.g., see \cite{yu,wang}. For instance,
in Ref. \cite{yu} the following spin-fermion representation of the
Hubbard operators
\begin{eqnarray}
X^{+0}=fS^{+}S^{-},\quad X^{0+}=f^{\dag}S^{+}S^{-},\quad
X^{-0}=fS^{-},\quad X^{0-}=f^{\dag}S^{+}
\label{eq:6.1}\end{eqnarray}
has been suggested,
which implies the identification
\begin{eqnarray}
&&|+\rangle\to |0_{Fermion};S^z=+1/2\rangle,\quad
|-\rangle\to |0_{Fermion};S^z=-1/2\rangle, \nonumber\\
&&|0\rangle\to |1_{Fermion};S^z=+1/2\rangle.
\label{eq:6.2}\end{eqnarray}
Note that all the Hubbard operators vanish on the state
$|1_{Fermion};S^z=-1/2\rangle$.

A similar map has been employed in Ref. \cite{wang}
except for a modification needed to explicitly recover the time-reversed
symmetry of the $t-J$ model. The latter is of no importance for us here,
so that later on we will keep referring to \cite{yu}, the more so, as
a trick suggested in \cite{wang} to consider an operator as a half-sum
of the same operator taken in two {\it different} representations
seems to pose a problem.

It can be easily checked that Eqs. (\ref{eq:6.1}) recover correctly
the algebra of the Hubbard operators within the subspace (\ref{eq:6.2}).
Nevertheless, operators $\vec S$ and $f$ cannot be considered as
those describing independently a local spin and a holon.
This results from the
$su(2|1)$ $(1/2,1/2)$ defining relation (\ref{eq:3.4}) which in terms of
(\ref{eq:6.1}) reads
\begin{eqnarray}
f^{\dag}f\,S^{+}S^{-}+ff^{\dag}=1.
\label{eq:6.3}\end{eqnarray}
This constraint is to be imposed in order to single out the three-dimensional
subspace (\ref{eq:6.2})
and plays the same role as Eq. (\ref{eq:2.77}) in the slave fermion
representation does.

To illustrate this point, consider the true fermionic
correlator
\begin{eqnarray}
G(t)=\langle f^{\dag}(t)f(0)\rangle\,\mid_{f^{\dag}f\,S^{+}S^{-}+ff^{\dag}=1}
= tr\,\{e^{-itH}f^{\dag}e^{itH}f\}\,\mid_{f^{\dag}f\,S^{+}S^{-}+ff^{\dag}=1}
\,,
\label{eq:6.4}\end{eqnarray}
where $H$ stands for any Hamiltonian that can be written in terms of Hubbard
operators.
On the other hand, consider
\begin{eqnarray}
\tilde G(t)=\langle f^{\dag}(t)f(0)\rangle
= tr\,\{e^{-itH}f^{\dag}e^{itH}f\}.
\end{eqnarray}
It is easily seen that
$$\tilde G(t)= G(t)+F(t),$$
where
$$F(t)=\langle 1_{Fermion};S^z=-1/2|\,e^{-itH}f^{\dag}e^{itH}f\,
|1_{Fermion};S^z=-1/2\rangle$$
$$= \langle 0_{Fermion};S^z=-1/2|\,e^{itH}\,|0_{Fermion};S^z=-1/2\rangle,$$
which means that the unphysical state $|1_{Fermion};S^z=-1/2\rangle$,
if not excluded by Eq. (\ref{eq:6.3}), makes a nontrivial contribution.

Though at half filling constraint (\ref{eq:6.3}) turns into identity
$1=1$, it is to be taken into consideration at any hole
concentration $\delta>0$. Otherwise, as the above mentioned example shows
unphysical states may affect the situation drastically.
In this regard, basic results of
Ref. \cite{yu}
where a motion of a hole has been investigated
in representation (\ref{eq:6.1}) without the constraint,
should have been revisited.

To look at all this from a viewpoint related to the path integral
(\ref{eq:4.1}-\ref{eq:4.2}),
consider the
$t$-dependent term in \cite{yu}:
\begin{eqnarray}
-t\sum_{ij}f_if^{\dag}_j\left[(\frac{1}{2}+S^z_i)S^{+}_j+
(\frac{1}{2}+S^z_j)S^{-}_i\right].
\label{eq:6.5}\end{eqnarray}
It is easily seen that the first term in Eq. (\ref{eq:4.3}) would just
correspond to this operator, provided one would consider $z$ and $\bar z$
to belong to $S^2$. But this is not the case and one must perform
the change (\ref{eq:4.tr}) first to bring the $SU(2|1)$ integral
to the $SU(2)$ form. As a result, one arrives at the first term of
Eq. (\ref{eq:4.8}) that on no account corresponds to (\ref{eq:6.5}).
All this amounts to saying that if one wrote down a path integral
for a partition function with the Hamiltonian (\ref{eq:6.5})
over
$D\mu_{SU(2)}D\mu_F$ and then performed the change of variables inverse to
(\ref{eq:4.tr}),
$$\xi\to\xi\sqrt{\frac{2q}{1+|z|^2}},\quad z\to \frac{z}
{\sqrt{1+\bar\xi\xi}},$$
then one would arrive at the representation (\ref{eq:4.1}) with a Hamiltonian
function describing a system quite distinct from the $t-J$ model.

For the sake of completeness, it should be also mentioned that
one can face an assertion that spin-charge degrees of freedom are separated
in certain instances, e.g., when $H_{t-J}$ is treated on a mean-field
level in a slave particle representation.
This assertion is correct, though has nothing to do with
Eq. (\ref{eq:4.eq}). The point is that the above-mentioned separation
holds for auxiliary fields related to electron spin and charge degrees
of freedom. Only some fixed combinations
of those fields can be associated with true spin variables.
In the slave fermion representation (\ref{eq:2.77})
one has
$$\frac{1}{2}(a^{\dag}a-b^{\dag}b)=S_z,\quad ab^{\dag}=S_{+},\quad
a^{\dag}b=S_{-},\quad a^{\dag}a+b^{\dag}b=2s\in N $$
whereas, for example,
operator $a$ taken in itself cannot be identified with a spin variable.
It would be therefore appropriate to refer this case to as
{\it a spinon}-charge separation rather than the spin-charge one.
Besides,
this separation breaks down beyond the mean-field approximation.

To conclude, we have shown that the $t-J$ Hamiltonian cannot be in general
reduced to that describing an interaction of two independent fields:
local $SU(2)$ spins and spinless fermions (holes), though this may occur
in some particular cases, e. g., in the linear spin wave
approximation. The consideration is based upon a crucial fact that the $t-J$
Hamiltonian can be embedded into a representation of the $su(2|1)$
superalgebra, which provides us with the $SU(2|1)$ path integral
representation of the partition function and, hence, with an effective total
action describing the system.

\vspace{1cm}
{\Large \bf VI. Acknowledgments}
\vspace{1cm}

The authors acknowledge the financial support
of the Russian Foundation for Fundamental Research
under grant No 96-01-00223.

\vspace{1cm}

{\Large \bf Appendix A}
\setcounter{equation}{0}
\def\theequation{A.\arabic{equation}}
\vspace{0.5cm}

To make the exposition self-contained and for the reader's convenience
we place in Appendices A, B and C some information concerning a definition
of the $su(2|1)$ superalgebra and its representations as well as of related
coherent states and
path integrals.

Appendix A serves to
recapitulate the necessary ingredients concerning the $su(2|1)$ superalgebra
and associated coherent states bearing in mind their
relevance for the $t-J$ model.

To visualize a route in general, we start with some preliminary remarks.
Given a Lie (super)algebra $g$ in an irreducible representation, one can
construct associated (super)coherent states $|(\cdotp)\rangle$ viewed as an
overcomplete basis in the corresponding representation of a (super)group
$G$, with $(\cdotp)$ specifying a point in the $G$-homogeneous
(super)manifold,
an orbit of $G$ in the coadjoint representation
\cite{perelomov,gradechi,topi}. Given further a Hamiltonian $H$ that appears
as an element of the $g$-enveloping algebra, one may evaluate a partition
function in the coherent-state basis, which naturally leads to a relevant
coherent-state
path integral \cite{kochetov3,kochetov4}.
The latter appears as a phase-space path integral
and provides a quantization of $H^{cl}$ on a coadjoint orbit. The crucial
point is that this quantization respects the underlying dynamical
(super)symmetry generated by $g$.

As is known, the $SU(2|1)$ supergroup in the fundamental representation
is the group of $(2+1)\times(2+1)$ unitary,
unimodular supermatrices with the Hermitian conjugate operation.
It is generated by even generators $\{B,Q_3,Q_{+},Q_{-}\}$ and
the odd ones $\{W_{+},W_{-},V_{+},V_{-}\}$ which
satisfy the following
commutation rules \cite{ritt}:
\begin{eqnarray}
[Q_3,Q_{\pm}]=\pm Q_{\pm},[Q_{+},Q_{-}]=2Q_3,[B,Q_{\pm}]=[B,Q_3]=0, \cr
[B,V_{\pm}]=\frac{1}{2}V_{\pm},[B,W_{\pm}]=-\frac{1}{2}W_{\pm},
[Q_3,V_{\pm}]=\pm\frac{1}{2}V_{\pm},[Q_3,W_{\pm}]=\pm\frac{1}{2}W_{\pm},\cr
[Q_{\pm},V_{\mp}]=V_{\pm},[Q_{\pm},W_{\mp}]=W_{\pm},
[Q_{\pm},V_{\pm}]=[Q_{\pm},W_{\pm}]=0,\cr
\{V_{\pm},V_{\pm}\}=\{V_{\pm},V_{\mp}\}=\{W_{\pm},W_{\pm}\}=
\{W_{\pm},W_{\mp}\}=0,\cr
\{V_{\pm},W_{\pm}\}=\pm Q_{\pm},\{V_{\pm},W_{\mp}\}=-Q_3\pm B.
\nonumber\end{eqnarray}

Let $|b,q,q_3\rangle$ stand for a vector of any abstract representation
of $su(2|1)$, where $b,q$ and $q_3$ denote the eigenvalues of the operators
$B,{\vec Q}^2$ and $Q_3$, respectively.
When considering the highest-weight state as the fiducial state $|0\rangle$,
the typical $SU(2|1)$ coherent state reads
\begin{eqnarray}
|z,\xi,\theta\rangle={\cal N}\,\exp( -\theta W_{-}
-\xi V_{-}+z Q_{-})|b,q,q\rangle,
\label{eq:a.1}\end{eqnarray}
where $(z,\xi,\theta)\in SU(2|1)/U(1)\times U(1)$.
We will be interested later on
in the so-called degenerate $b=q$ representation
which happens to be relevant for the $t-J$ model. It is specified by
\begin{eqnarray}
W_{-}|q,q,q\rangle=0
\label{eq:a.2}\end{eqnarray}
and is called the $(q,q)$ representation
with the dimension $4q+1$ \cite{ritt}.
This representation is spanned by $2q+1$ vectors
$\{|q,q,m\rangle,\quad -q\leq m\leq q\}$ of the even (bosonic) sector
and $2q$ vectors $\{|q+1/2,q-1/2,m\rangle,\quad -q+1/2\leq m\leq q-1/2\}$
that correspond to the odd (fermionic) one.
Both the second and third order Casimir operators are zero in this
representation.
The coherent state (\ref{eq:a.1}) is reduced in the $(q,q)$ representation to
\begin{eqnarray}
|z,\xi\rangle=(1+|z|^2+\bar\xi\xi)^{-q}e^{-\xi V_{-}+z Q_{-}}|q,q,q\rangle,
\label{eq:a.3}\end{eqnarray}
wherein we have evaluated the normalization factor explicitly.

\vspace{1cm}

{\Large \bf Appendix B}
\setcounter{equation}{0}
\def\theequation{B.\arabic{equation}}
\vspace{0.5cm}
\vspace{1cm}

In this Appendix we describe the $SU(2)$ and standard "fermionic"
path integrals, which is necessary to interpret properly
the transformed $SU(2|1)$ path integral (\ref{eq:4.7}).

Consider the $SU(2)$ algebra
$$[S_z,S_{\pm}]=\pm S_{\pm},\quad [S_{+},S_{-}]=2S_z.$$
Corresponding coherent states for the UIR with $s\in N/2,\quad
\vec S^2=s(s+1)$ are given by
$$|z\rangle_{SU(2)}=(1+|z|^2)^{-s}e^{zS_{+}}|s,-s\rangle,$$
where $z\in SU(2)/U(1)=S^2$, and $S_z|s,m\rangle=m|s,m\rangle$.
Given a Hamiltonian $H=H(\vec S)$, the partition function reads
\begin{eqnarray}
Z_{SU(2)}=\int\limits_{z(0)=z(\beta)}D\mu_{SU(2)}(\bar z,z)
\exp[{\cal A}(\bar z,z)].
\label{eq:a1}\end{eqnarray}
where an effective $SU(2)$ action
\begin{eqnarray}
{\cal A}= s\int_0^{\beta}\frac{\dot{\bar z}z-\bar z\dot z}{1+|z|^2}dt-
\int_0^{\beta}H^{cl}(\bar z,z)dt,
\label{eq:a2}\end{eqnarray}
and  $H^{cl}=\langle z|H|z\rangle$.
The classical counterparts of the $SU(2)$ generators are easily found to be
\begin{eqnarray}
S_z^{cl}=-s(1-|z|^2)w^{(0)},\quad (S_{+})^{cl}=2szw^{(0)},\quad
(S_{-})^{cl}=2s\bar zw^{(0)},
\label{eq:a3}\end{eqnarray}
where $w^{(0)}=(1+|z|^2)^{-1}.$
In Eq. (\ref{eq:a1}) $D\mu_{SU(2)}(\bar z,z)$ stands for the infinite
pointwise product of the $SU(2)$ invariant measures,
\begin{eqnarray}
D\mu_{SU(2)}=\prod_j^{\infty}\frac{2s+1}{2\pi i}\frac{d\bar z_jdz_j}
{(1+|z_j|^2)^2}.
\label{eq:a4}\end{eqnarray}
For more details see Ref. \cite{kochetov1}.

The $SU(2)$ action on $S^2$ reads
\begin{equation}
z\to gz=\frac{uz+v}{-\bar vz+\bar u},
\label{eq:a5}\end{equation}
where
\begin{displaymath}
{\left(\begin{array}{cc}
u & v \\
{-\bar v} & {\bar u}\\
\end{array}\right) =g\in SU(2)}.
\end{displaymath}

As the second example consider the more familiar fermionic oscillator
algebra generated by
$$f,\quad f^{\dag},\quad f^{\dag}f,\quad I,$$
with $\{f,f^{\dag}\}=1$. The corresponding coherent states
$$|\xi\rangle_F=(1+\bar\xi\xi)^{-1/2}e^{\xi f^{\dag}}\,|0\rangle_{Fock}$$
are parametrized by the generators of the Grassmann algebra
$Q_0\oplus Q_1$:
$$\{\xi,\bar\xi\}=0,\quad \xi^2=\bar \xi^2=0,\quad
\xi,\bar\xi\in Q_1,\quad 1,\,\bar\xi\xi\in Q_0.$$
These states give rise to the following representation
\begin{eqnarray}
Z_F=\int\limits_{\xi(0)=-\bar\xi(\beta)}D\mu_F(\bar\xi,\xi)
\exp[{\cal A}(\bar\xi,\xi)],
\label{eq:a6}\end{eqnarray}
where
\begin{eqnarray}
{\cal A}=\frac{1}{2}\int_0^{\beta}(\dot{\bar\xi}\xi-\bar\xi\dot\xi)dt-
\int_0^{\beta}H^{cl}(\bar\xi,\xi)dt,\quad H^{cl}=\langle\xi|H|\xi\rangle,
\label{eq:a7}\end{eqnarray}
with $f^{cl}=\xi,\,(f^{\dag})^{cl}=\bar\xi$ and $(f^{\dag}f)^{cl}=\bar\xi\xi$.
Here $D\mu_F$ stands for
\begin{eqnarray}
D\mu_F=\prod_j^{\infty}d\bar\xi_jd\xi_j
\label{eq:a8}\end{eqnarray}
that is clearly invariant with respect to
a shift by a Grassmann parameter,
$\xi\to\xi+\xi_0,$ combined with the phase transformation,
$\xi\to\xi e^{i\alpha},\quad \alpha\in Q_0$.

\newpage

{\Large \bf Appendix C}
\setcounter{equation}{0}
\def\theequation{C.\arabic{equation}}
\vspace{0.5cm}

\vspace{1cm}

Suppose we are given two objects: a two-dimensional sphere $S^2$ with local
complex coordinates $\bar z,z$ and a
Grassmann algebra with two generators $\theta$ and $\bar\theta$ and with a
generic element
$$f(\theta,\bar\theta)=f_0+f_1\theta+f_2\bar\theta+f_3\bar\theta\theta,$$
where $f_i$ are complex numbers.
The $N=2$ supermanifold $S^{2|2}$ is the pair $(S^2,{\cal A}_{S^2})$ where
$S^2$ is the two-sphere and ${\cal A}_{S^2}$ is a sheaf of supercommutative
(Grassmann valued) algebras on $S^2$ with a general section (element)
\begin{eqnarray}
h(\bar z,z;\bar\theta,\theta)=h_0(\bar z,z)+\theta h_1(\bar z,z)+\bar\theta
h_2(\bar z,z)+\bar\theta\theta h_3(\bar z,z),
\label{eq:b.1}\end{eqnarray}
where $h_i$ belong to $C^{\infty}(S^2)$.
We follow here a general defenition of a supermanifold given by Berezin
\cite{berezin1} (see also Refs. \cite{gradechi}).

The pair $(z,\theta)$ serves as supercoordinates on $S^{2|2}$. The very
same role, however, can be played by any set of even and odd
generators of ${\cal A}_{S^2}$, provided Eq. (\ref{eq:b.1})
still holds in new variables. We are interested in a reparametrization of the
specific type,
\begin{eqnarray}
w=w(z;\bar\theta,\theta),\quad \bar w=\bar w(\bar z;\bar\theta,\theta),\quad
\xi=\xi(\bar z,z;\theta),\quad \bar\xi=\bar\xi(\bar z,z;\bar\theta).
\label{eq:b.2}\end{eqnarray}
Under some restrictions on functions $w(;), \ldots, \bar\xi(;)$ Eq.
(\ref{eq:b.2}) introduces a new set of coordimates on $S^{2|2}$.
The most important requirement
for us is that the map $(\bar z,z)\to spec\, (\bar w,w)$ is to be a
diffeomorphism $S^2 \to S^2$, where a spectrum of any element of
the Grassmann  algebra is defined by
$$spec\, f=f\mid_{\bar\theta=\theta=0}=f_0\in C.$$

Now we are in a position to define an integration on $S^{2|2}$. For a function
$F(\bar w,w;\bar\xi.\xi)$ we have by definition \cite{berezin1}
\begin{eqnarray}
&&\int_{S^{2|2}}F(\bar w,w;\bar\xi,\xi) \rho(\bar w,w;\bar\xi,\xi)
d\bar wdwd\bar\xi d\xi=\nonumber \\
&&\int\limits_{\bar v,v\in spec (\bar w,w)=S^2}F(\bar v,v;\bar\xi,\xi)
\rho(\bar v,v;\bar\xi,\xi)d\bar vdvd\bar\xi d\xi,
\label{eq:b.3}\end{eqnarray}
where $\rho(\bar w,w;\bar\xi,\xi)d\bar wdwd\bar\xi d\xi$ is the $SU(2|1)$
invariant volume element (\ref{eq:2.4}) and
$v=w(z,o)$ and $\bar v=\bar w(\bar z,o)$ are diffeomorphisms $S^2\to
S^2$. Here an integration over $d\bar vdv$ is understood in a usual manner,
whereas an integration over $d\bar\xi d\xi$ is to be carried out in accordance
with the Berezin' convention. The last point to be noted is that any change of
variables in the lhs of Eq. (\ref{eq:b.3}) gives rise to a superdeterminant
(Berezian) of a corresponding transformation matrix.


\newpage
\begin{thebibliography}{99}
\bibitem{and} P.W. Anderson, Science {\bf 235}, 1196 (1987)
\bibitem{zhang} F.C. Zhang and T.M. Rice, Phys. Rev. {\bf B37}, 3759 (1988)
\bibitem{kane} C.L. Kane, P.A. Lee and N. Read, Phys. Rev. {\bf B39},
6880 (1989)
\bibitem{lee} P.A. Lee, Phys. Rev. Lett. {\bf 63}, 680 (1989);
P.A. Lee and N. Nagaosa, Phys. Rev. {\bf B46}, 5621 (1992)
\bibitem{yu} J.L. Richard and V.Yu. Yushankhai, Phys. Rev. {\bf B47}, 1103
(1993)
\bibitem{wang} Y.R. Wang and M.J. Rice, Phys. Rev {\bf B49}, 4360 (1994)
\bibitem{wig} P.B. Wiegmann, Phys. Rev. Lett. {\bf 60}, 821 (1988)
\bibitem{perelomov} A. Perelomov, "Generalized coherent states and their
applications" (Springer-Verlag, Berlin, 1986);
A.B. Balantekin, H.A. Schmitt and B.R. Barrett, J. Math.
Phys. {\bf 29}, 1634 (1988);
A.B. Balantekin, H.A. Schmitt and P. Halse, J. Math. Phys.
{\bf 30}, 274 (1989)
\bibitem{gradechi} A.M. El Gradechi, J. Math. Phys. {\bf 34}, 5951 (1993);
"Geometric Quantization of an $Osp(1|2)$ Orbit" Preprint CRM-1882 (1994);
A.M. El Gradechi and L.M. Nieto,
"Supercoherent states, Super K\"ahler Geometry and Geometric Quantization"
Preprint CRM-1876 (1994)
\bibitem{topi} A. Pelizzola and C. Topi, Int. J. Mod. Phys. {\bf B5} (1991)
3073
\bibitem{kochetov2} E.A. Kochetov, Phys. Rev. {\bf B52}, 4402 (1995)
\bibitem{kochetov5} E.A. Kochetov, V.S. Yarunin and M.E. Zhuravlev,
to be published
\bibitem{hubbard} J. Hubbard, Proc. R. Soc. London {\bf A285}, 542 (1965)
\bibitem{kochetov4} E.A. Kochetov, Phys. Lett. {\bf A217}, 65 (1996);
see also E.A. Kochetov, J. Phys. {\bf A26}, 3489 (1993)
\bibitem{bars} I. Bars and M. G\"unaydin, Commun. math. Phys. {\bf 91}, 31
(1983)
\bibitem{berezin2} F.A. Berezin, Commun. math. Phys. {\bf 40}, 153 (1975)
\bibitem{anderson} P.W. Anderson, Phys. Rev {\bf 86}, 694 (1952)
\bibitem{kubo} R. Kubo, Phys. Rev {\bf 87}, 568 (1952)
\bibitem{horsch} G. Martinez and P. Horsch, Phys. Rev {\bf B44}, 317 (1991)
\bibitem{kochetov3} E.A. Kochetov, J. Math. Phys. {\bf 36}, 1666 (1995)
\bibitem{ritt} M. Scheunert, W. Nahm and V. Rittenberg,
J. Math. Phys. {\bf 18}, 155 (1977)
\bibitem{kochetov1} E.A. Kochetov, J. Math. Phys. {\bf 36}, 4667 (1995)
\bibitem{berezin1} F.A. Berezin, "Introduction to Superanalysis"
(Reidel, Dordrecht, 1987)

\end{thebibliography}
\end{document}